# Large Rashba spin-orbit coupling in metallic SrTaO$_3$ thin films

Hikaru Okuma[1]*, Yumiko Katayama[1], Fukunobu Kadowaki[1], Yuki Tokumoto[2], and Kazunori Ueno[1]

1 *Graduate School of Arts and Sciences, The University of Tokyo, Komaba, Meguro-ku, Tokyo 153-8902, Japan.*

2 *Institute of Industrial Science, The University of Tokyo, Komaba, Meguro-ku, Tokyo 153-8505, Japan.*

*Author to whom correspondence should be addressed: hikaruokuma613@gmail.com

## Abstract

Epitaxial thin films of SrTaO$_3$ with thickness ($t$) smaller than 74 nm were successfully fabricated on an insulator (LaAlO$_3$)$_{0.3}$(Sr$_2$AlTaO$_6$)$_{0.7}$ substrate. Films with $t$ above 8.6 nm showed metallic conduction. Both conductivity and a mobility showed a decrease with increasing $t$ above 42 nm, suggesting the instability of thick SrTaO$_3$ films. This instability was also supported by TEM image and XRD intensity. For the metallic films with $t$ below 25 nm, energy band splitting due to spin-orbit coupling ($\Delta_{so}$) and Rashba parameter ($\alpha_R$) were deduced from an analysis of a magnetoresistance using two-dimensional weak antilocalization theory. The values of $\Delta_{so}$ ranged from 26 to 120 meV, which were the largest among other metallic oxide films, such as SrNbO$_3$, SrIrO$_3$, and La$_{2/3}$Sr$_{1/3}$MnO$_3$ thin films, indicating that spin-orbit coupling in SrTaO$_3$ was the largest among the metallic perovskite oxides reported so far. The values of $\alpha_R$ for our SrTaO$_3$ films ranged from 8.8×10$^{-13}$ to 1.7×10$^{-12}$ eV m, which were much larger than those reported for other metallic oxide thin films.

For conductive 4$d$- and 5$d$-electron ABO$_3$ materials with strong spin-orbit coupling (SOC), interesting physical properties, such as a topological Hall effect[1-3] and Dirac semimetal states[4,5], were found to emerge. A two-dimensional (2D) electron gas on a semiconductor KTaO$_3$ (KTO) with the heavy element of Ta showed many fascinating phenomena[6,7], such as large Rashba splitting[8,9] and various topological physical properties[10,11]. Furthermore, an interface on KTO showed a higher superconducting transition temperature[12] than an interface on SrTiO$_3$[13]. We have reported strong SOC in metallic SrNbO$_3$ (SNO) films with a 4$d$-electron[14]. However, except for SrIrO$_3$ (SIO), a metallic ABO$_3$ material with a 5$d$-electron has been rarely investigated despite the expectation that it would have strong SOC. Therefore, the fabrication of new Ta-based metallic oxides ABO$_3$ is strongly desired.

Density functional theory calculations predict that SrTaO$_3$ exhibits metallic conduction and has a lattice constant of around $a$=0.4007 nm[15], which is larger than the lattice constant of SrTiO$_3$ ($a$=0.3905 nm). To our knowledge, however, neither the crystal structure nor electronic properties of the SrTaO$_3$ thin films have been elucidated experimentally. Therefore, it is important to first clarify whether epitaxial thin films of SrTaO$_3$ can be indeed fabricated on a substrate where a large lattice mismatch is present. It has been reported quite recently that around the interface between SrTaO$_3$ and oxide semiconductors such as SrTiO$_3$, KTO, BaSnO$_3$, and SrSnO$_3$ not only SrTaO$_3$ but also oxide semiconductors near the interface could contribute to conduction[16-19].

In this work, we fabricated SrTaO$_3$ films with various thicknesses ($t$). We employed an insulator (LaAlO$_3$)$_{0.3}$(Sr$_2$AlTaO$_6$)$_{0.7}$ (LSAT) substrate ($a$ = 0.3868 nm) to exclude interface conduction. Interestingly, epitaxially grown single crystal films were successfully obtained although a large mismatch between the substrate and the films is present. The thin films with $t$ ranging from 8.6 to 25 nm showed a metallic behavior, and the metal-insulator transition occurred by reducing $t$. Magnetoresistance (MR) was examined on the metallic films, and we estimated Rashba SOC on SrTaO$_3$ with an analysis of MR using weak antilocalization (WAL) theory.

The SrTaO$_3$ thin films were fabricated on the (001)-oriented LSAT substrate by a pulsed laser deposition method in a vacuum with a back pressure of 10$^{-7}$ Torr at 700-900 °C. A ceramic target composed of Sr$_2$Ta$_2$O$_7$ was ablated by a KrF excimer laser ($\lambda$ = 248 nm) with a repetition rate of 5 Hz and an energy fluence of 0.98-1.08 J/cm$^2$. The growth of the SrTaO$_3$ films was examined by cross-section transmission electron microscopy (TEM) and an out-of-plane scan and reciprocal space mapping (RSM) with X-ray diffraction (XRD) (Rigaku, Smart Lab). The thickness of the films was directly measured by a surface profiler (Bruker, Dektak XT) and checked by TEM for selected ones. In addition, the surface uniformity was estimated with an atomic force microscope (AFM). Aluminum wires were ultrasonically bonded on a rectangular sample, working as current and voltage probes. Ohmic contacts were confirmed down to the lowest temperature ($T$). Longitudinal resistivity ($\rho$) and Hall resistance ($R_{yx}$) were measured using the van der Pauw configuration in a physical properties measurement system (Quantum Design, PPMS) from 2 to 300 K and –7 to 7 T.

Figure 1(a) shows a cross-sectional TEM image of the SrTaO$_3$ film grown on the LSAT substrate. Both the interface and the surface were flat. Surface flatness was also confirmed by an AFM image (see Fig. 1(g)). The film thickness estimated from the TEM image was $t = 72\pm1$ nm, which is in agreement with $t = 74 \pm3$ nm measured by the surface profiler. Figure 1(b) shows a magnified TEM image at the interfacial region between the substrate and the film. Epitaxial growth of the film was observed. Figure 1(c) depicts Fast Fourier Transform (FFT) patterns of the TEM image at the interface between the film and the substrate. Reciprocal lattices for the film and the substrate were clearly observed. In-plane lattice constants ($a$) and out-of-plane lattice constants ($c$) of the film were estimated to be $a = 0.405\pm0.003$ nm and $c = 0.405\pm0.003$ nm, respectively. Figures S1(c) and S1(d) show a TEM image and an FFT pattern near the film surface, respectively. A halo pattern in the FFT pattern indicates the existence of amorphous or polycrystalline domains near the surface.

Figure 1(d) shows an out-of-plane $2\theta/\omega$ scan of the XRD for SrTaO$_3$/LSAT with $t = 74$ nm. The value of $c$ for the SrTaO$_3$ film was extracted from the SrTaO$_3$(002) peak. RSM was performed on a (103) reflection of the same sample, as shown in Fig. 1(e), where $Q_x$ and $Q_z$ denote reciprocal space coordinates along the $a$- and $c$-axis, respectively. The reciprocal lattice point for the film calculated from the FFT patterns in Fig.1 (c) is marked with a black cross, which is located in the high-intensity area elongated along the $Q_x$ axis. Lattice constants $a$ and $c$ for films with various $t$ were extracted from RSM and $2\theta/\omega$ scan, respectively, as shown in Figs. S2(a) and S2(b). $a$ and $c$ were $0.402\pm0.002$ nm and $0.406\pm0.002$ nm, respectively, which were close to the lattice constant predicted by calculation ($a=c = 0.401$ nm)[15].

To examine the variations in crystallinity as a function of $t$, rocking curves of the SrTaO$_3$ (002) reflection peaks for the SrTaO$_3$/LSAT were measured. The full width at half maximum (FWHM) of the rocking curve was extracted for each film. As shown in Fig. 1(f), the FWHM decreased with increasing $t$ from 12 to 38 nm. This can be explained by a large number of dislocations or defects near the interface areas. With increasing $t$ above 38 nm, the FWHM unexpectedly increased, indicating degradation of crystalline quality. This was also observed from the $t$ dependence of the integrated intensity of the film in the $2\theta/\omega$ XRD pattern, as shown in Fig. 1(f). Assuming that the crystal quality remains unchanged, the XRD intensity is proportional to $t$. The integrated intensity was linearly increased with $t$ up to 25 nm, but decreased with a further increase in $t$ above 25 nm. This indicates that the volume fraction of the (001)-oriented film decreased with increasing $t$, suggesting the formation of a polycrystalline or amorphous film. This degradation in the thick film is also consistent with the observation of the halo FFT pattern for the thick film with $t = 74$ nm.

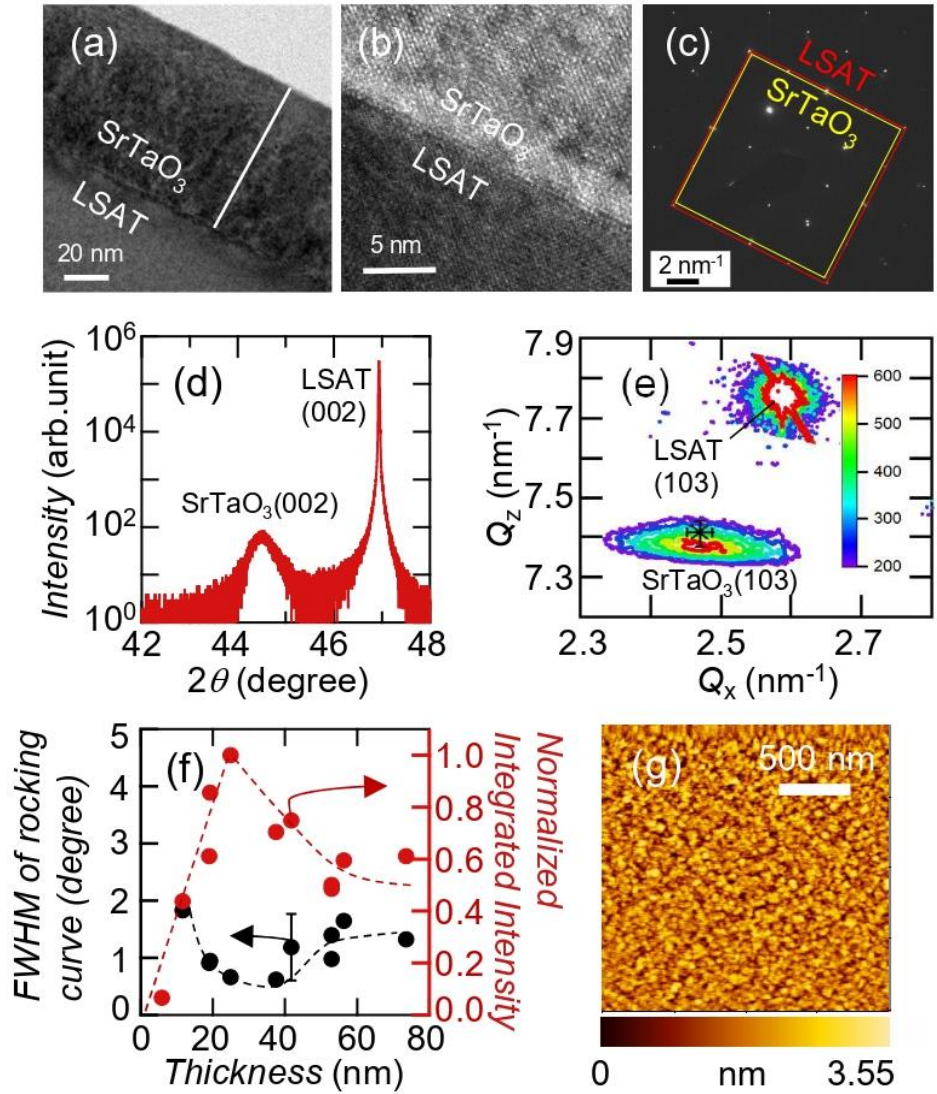

FIG. 1. (a,b) Cross-sectional TEM images for a SrTaO3 film with $t$ = 74 nm grown on the LSAT substrate. The interfacial region between the substrate and the film is shown in (b). (c) FFT patterns of the TEM image around the interface and the substrate. The yellow and red boxes represent reciprocal lattices for the film and the substrate, respectively. (d) Out-of-plane $2\theta/\omega$ scan of XRD for SrTaO3/LSAT with $t$ = 74 nm. (e) RSM on a (103) reflection of SrTaO3/LSAT with $t$ = 74 nm, where $Q_x$ and $Q_z$ denote reciprocal space coordinates along the $a$- and $c$-axis, respectively. A black cross corresponds to a reciprocal lattice point for the film calculated from the FFT pattern shown with the yellow box in (c). (f) (left) FWHM of the rocking curve versus $t$ plot. (right) Integrated intensity for the SrTaO3(002) peak of the out-of-plane $2\theta/\omega$ scan normalized by the intensity for the LSAT substrate plotted against $t$. (g) AFM surface morphology of the film with $t$ = 74 nm.

Figure 2(a) shows $\rho(T)$ for the SrTaO$_3$ films with $t$ ranging from 5.8 to 74 nm. At room temperature, $\rho(T)$ decreased with increasing $t$ up to 25 nm and then increased. For the thinnest film with $t$ = 5.8 nm, the sheet resistance ($R_s$) at room temperature $R_s$(300K) is 29.7 kΩ, which was larger than the quantum resistance $R_Q$ = 25.8 kΩ, and $\rho(T)$ showed an abrupt increase at around $T$ = 0. For all other films with $t \geq$ 8.6 nm, $R_s(T)$ is always smaller than $R_Q$. For the films with $t$ = 12-42 nm, $\rho(T)$ shows a decrease with decreasing temperature from 300 K and shows a resistivity minimum at a certain temperature ($T_{min}$), as indicated by an arrow in Fig. 2(a). At temperatures above $T_{min}$, $\rho$ is proportional to $T^2$, which originates from a Fermi-liquid behavior of quasiparticles caused by electron correlation. The increase in $\rho$ upon cooling is explained by weak localization (WL). WL at low $T$ and the crossover to the Fermi-liquid behavior at high $T$ are commonly observed in 2D disordered systems such as SrVO$_3$ (SVO)[20], SNO[14,21], SrRuO$_3$[22], and SIO[23]. In the weakly localized regime, resistivity increases at low temperature due to electron interference. In a 2D system, the quantum correction to the sheet conductance ($\sigma_S=1/R_s$) is expressed as $\Delta\sigma_S(T) = a \ln T$[24,25]. Here, the coefficient $a$ is of the order of the quantum conductance $e^2/\pi h$, where $e$ is the elementary charge and $h$ is the Planck constant. As shown in Fig. S3, $\sigma_S(T)$ for the 25 nm-thick film followed the ln$T$ dependence over substantial $T$ ranges. We will discuss transport properties in the weakly localized regime in detail later.

Next, we focus on the $t$ dependence of transport properties such as carrier density ($n$) and carrier mobility ($\mu$), which were estimated from Hall measurements. Except for the thinnest film with $t$ = 5.8 nm, whose $R_{yx}$ was below the experimental sensitivity, a negative linear $R_{yx}$ was observed for all films, indicating a single electron band. Figure 2(b) shows $\rho$, $\mu$, and $n$ for the films shown in Fig. 2(a) measured at 2 and 200 K. For all films, both $\mu$ and $n$ are nearly independent of temperature. With increasing $t$ from 8.6 to 19 nm, $\mu$ shows an increase and a saturation. Then, $\mu$ shows a decrease with a further increase in $t$ from 42 nm. $n$ also showed similar $t$ dependence. The low $\mu$ in a thinner film indicates an enhancement of disorder effects at around the interface and the surface, such as strain effects and oxygen vacancies[26-28]. These disorders bring the system close to the metal-insulator transition. It is likely that the density of states decreased as the electrons are gradually localized, and then the carrier density is also reduced in the thinner film. In a thicker film above 25 nm, the degradation in the crystallinity probably reduced both $n$ and $\mu$. As discussed above, the integrated intensity of SrTaO$_3$ (200) diffraction in the XRD showed a decrease in a thicker film above 25 nm, probably due to the formation of other phases than (200) oriented single crystal. These other phases worked as scattering center and reduced $\mu$. In addition, the carrier density in these other phases is probably smaller than in the metallic SrTaO$_3$ single crystal, and then reduced $n$. In short, all these results consistently show that the single crystal structure of SrTaO$_3$ is stable only for thin films with $t \leq$ 25 nm, while it becomes abruptly unstable for thicker films with $t >$ 25 nm.

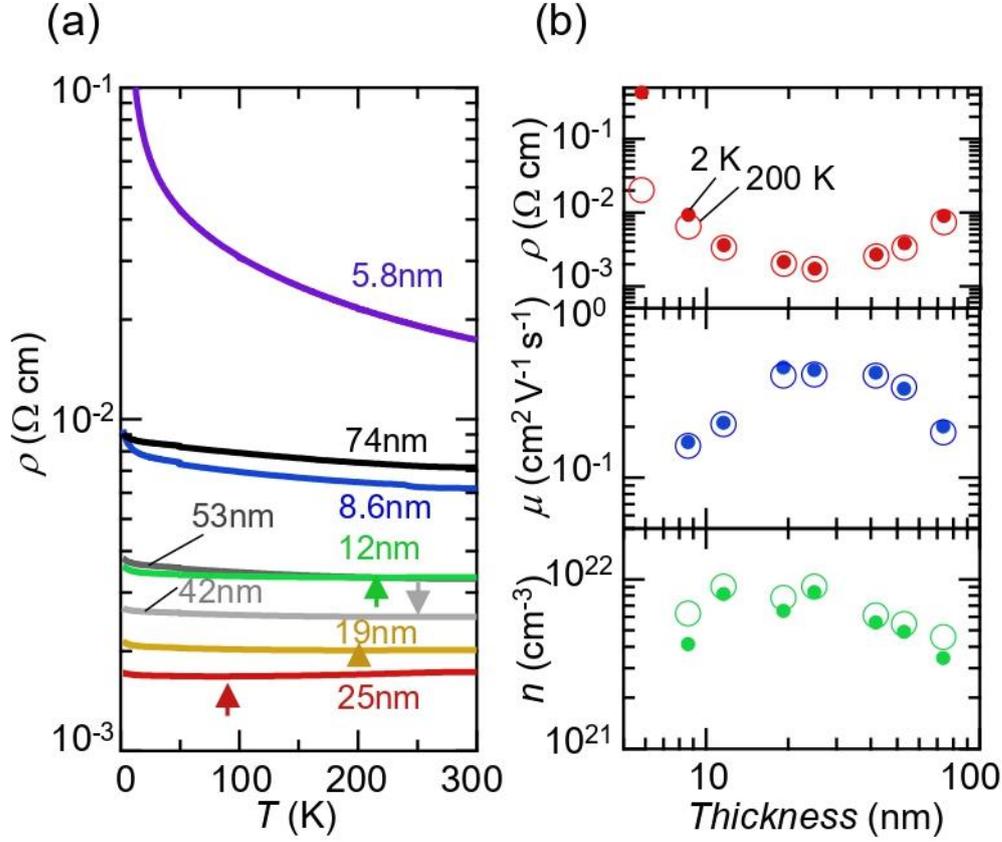

FIG. 2. (a) Temperature dependence of the longitudinal resistivity ($\rho$) for the SrTaO$_3$/LSAT films with various $t$ values ranging from 5.8 to 74 nm. Arrows indicate temperatures for the resistivity minima. (b) $\rho$, $\mu$ (electron mobility), and $n$ (electron density) at 2 K (solid circles) and 200 K (open circles) plotted against $t$.

Finally, we examined the Rashba SOC in SrTaO$_3$ with the transport properties in the weakly localized regime. Figure 3(a) shows a normalized MR, which is defined as $100 \times [\rho(B) - \rho(0)]/\rho(0)$, where $B$ is a perpendicular magnetic field, at 2 K for films with $t$ ranging 8.6-25 nm. For the system with SOC, when a weak $B$ is applied, the spin phase shift is compensated by the vector potential, and the conductivity decreases. Therefore, the WAL contribution can be observed at low $B$[14,29]. The WAL effect is usually observed in systems with strong SOC, such as semiconductors[30-34], heavy metals[35-39], and layered materials[40-44]. In some of these systems, a very short spin relaxation length was reported. To estimate the spin relaxation length for the SrTaO$_3$ films, we fit the data of the magnetoconductance using the Iordanskii-Lyanda-Pikus (ILP) theory[45,46] for a 2D electronic system, which is expressed as:

$$\frac{\Delta \sigma_s(B)}{\sigma_0} = -\left[\frac{1}{2}\Psi\left(\frac{1}{2}+\frac{B_{in}}{B}\right) - \frac{1}{2}\ln\frac{B_{in}}{B} - \Psi\left(\frac{1}{2}+\frac{B_{in}+B_{so}}{B}\right) + \ln\frac{B_{in}+B_{so}}{B}\right.$$
$$\left. - \frac{1}{2}\Psi\left(\frac{1}{2}+\frac{B_{in}+2B_{so}}{B}\right) + \frac{1}{2}\ln\frac{B_{in}+2B_{so}}{B}\right], \quad (1)$$

where $\sigma_0 = e^2/\pi h$, $\Psi$ is the digamma function, and $B_{in} = \frac{\hbar}{4eL^2_{in}}$ and $B_{so} = \frac{\hbar}{4eL^2_{so}}$ are effective fields related to inelastic and spin-orbit scattering, respectively. Here, $L_{in}$ and $L_{so}$ are the inelastic and spin relaxation lengths, respectively, and $\hbar$ is the Dirac's constant. Figure 3(b) shows $\Delta\sigma_S(B) = \sigma_S(B) - \sigma_S(0)$ in units of $e^2/\pi h$ at 2 K for films with $t$ ranging from 8.6 to 25 nm and a fit of each data to the ILP theory (Eq. (1)). $\Delta\sigma_S(B)$ well followed Eq. (1), and we deduced $L_{so}$ and $L_{in}$ from each fit. $L_{in}$ is larger than the thickness for the films with $t$ below 25 nm, indicating 2D nature of these films (see Fig. S4). Figure 3(c) shows a $t$ dependence of $L_{so}$ and comparison with those for other metallic oxide films: SNO[14], SIO[5,47], SIO/LaMnO$_3$ (LMO)[48]. It is commonly observed for a series of the SrTaO$_3$, SNO, and SIO films that $L_{so}$ is large in the small $t$ region, where the inelastic scattering length is comparable to the spin-orbit scattering length, but it decreases with increasing $t$, and finally shows a trend to saturate in the large $t$ region. Notably, the values of $L_{so}$ obtained in the SrTaO$_3$ films with $t$ = 12-25 nm are at least as small as or slightly smaller than the values reported for other metallic oxide films, indicating that SOC in the SrTaO$_3$ films is the largest among the metallic perovskite oxide films reported so far. Furthermore, the values of $L_{so}$ for the SrTaO$_3$ films ($L_{so}$ = 9-17 nm) are smaller than or comparable to the values reported for layered materials (MoS$_2$ with $L_{so}$ = 100-500 nm[41], TaSe$_2$ with $L_{so}$ = 17 nm[42], VSe$_2$ with $L_{so}$ = 17 nm[43], Pb(Bi$_{1-x}$Sb$_x$)$_2$(Te$_{1-y}$Se$_y$)$_4$ with $L_{so}$ = 13 nm[44]), semiconductors (KTO with $L_{so}$ = 20-60 nm[31], SrTiO$_3$ with $L_{so}$ = 55-90 nm[33], GaSe with $L_{so}$ = 14-18 nm[34]), and heavy metals (Au with $L_{so}$ = 11 nm[36], Pt with $L_{so}$ = 3.0-9.5 nm[38,39]).

Figures 4(a) and 4(b) show the energy band splitting due to SOC ($\Delta_{so}$) and Rashba parameter ($\alpha_R$) as a function of $R_s$, respectively, calculated with the following relation, assuming a parabolic band structure and a single spin winding in the spin texture:

$$L_{so} = \frac{\hbar^2}{2m\alpha_R}, \quad (2)$$
$$\Delta_{so} = 2k_F\alpha_R, \quad (3)$$
$$k_F = \sqrt{2\pi n_s}, \quad (4)$$

where $m$ is the effective mass, $k_F$ is the Fermi wave number in circular Fermi surface of a 2D system, and $n_s = nt$ is the sheet carrier density. We employed $m = 2.8m_0$ reported for SNO[49], where $m_0$ is the free electron mass. Thus, the values of $\Delta_{so}$ ranged from 26 to 120 meV, which were the largest among the oxide materials. The values of $\alpha_R$ ranged from $8.8\times10^{-13}$ to $1.7\times10^{-12}$ eV m, which were also much larger than the values reported for thin films of metallic oxides such as SIO[50], SIO/LMO[48] and La$_{2/3}$Sr$_{1/3}$MnO$_3$[51], and comparable to the values reported for the heterointerface (LaAlO$_3$/SrTiO$_3$)[32]. It is noted that the values of $\alpha_R$ for SrTaO$_3$ were slightly larger than the values for SNO films[14], which were obtained assuming that $m$ in SrTaO$_3$ is equal to that in SNO. Since the bandwidth of $d$ orbit for heavy element is broader than that for light element (e.g., the larger bandwidth of Nb 4$d$ for SNO than that of V 3$d$ for SrVO$_3$ was reported by DFT calculation[49]), $m$ in SrTaO$_3$ would be smaller than that in SNO. Then, $\alpha_R$ for the SrTaO$_3$ films is expected to be larger than that estimated here.

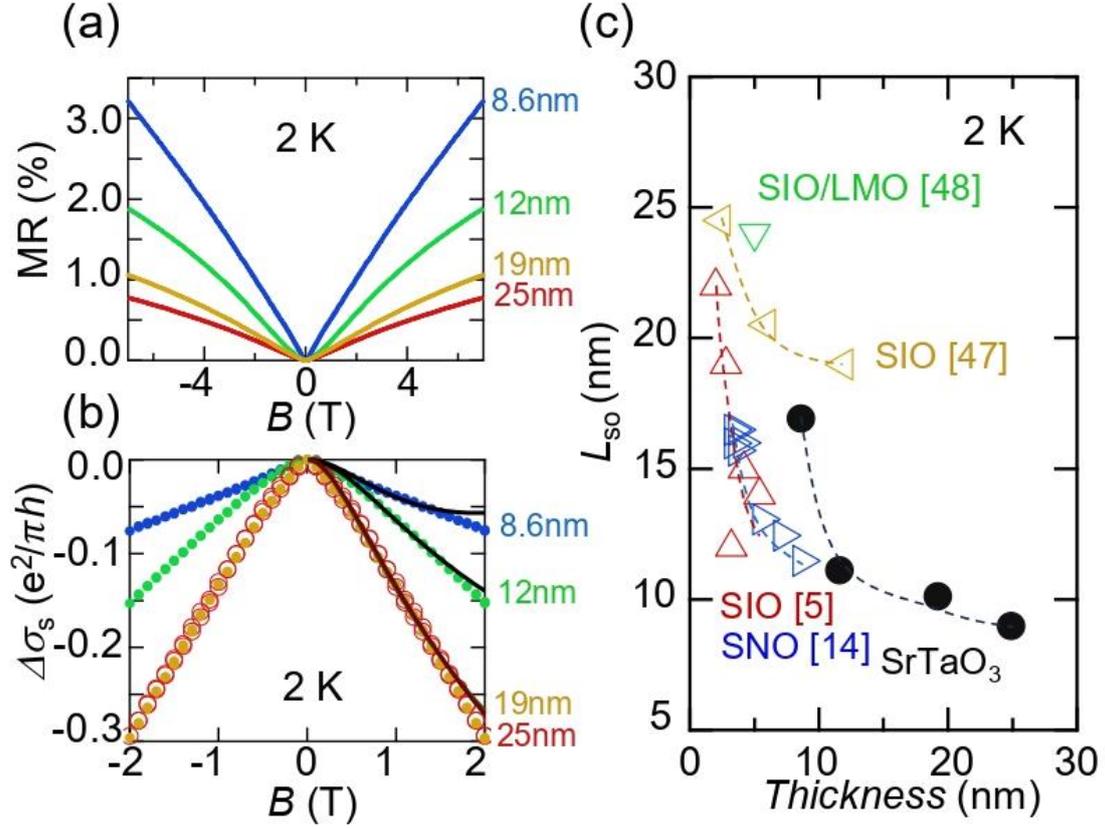

FIG. 3. (a) Normalized MR and (b) $\Delta\sigma_s$ in units of e$^2$/$\pi h$ measured in the perpendicular magnetic field at 2K for films with $t$ ranging 8.6-25 nm. The solid lines are fits to the data based on the ILP theory [Eq. (1)]. (c) Spin relaxation length for metallic oxide materials, including SrTaO$_3$ (this study), SNO[14], SIO[5,47], SIO/LMO[48] plotted against $t$.

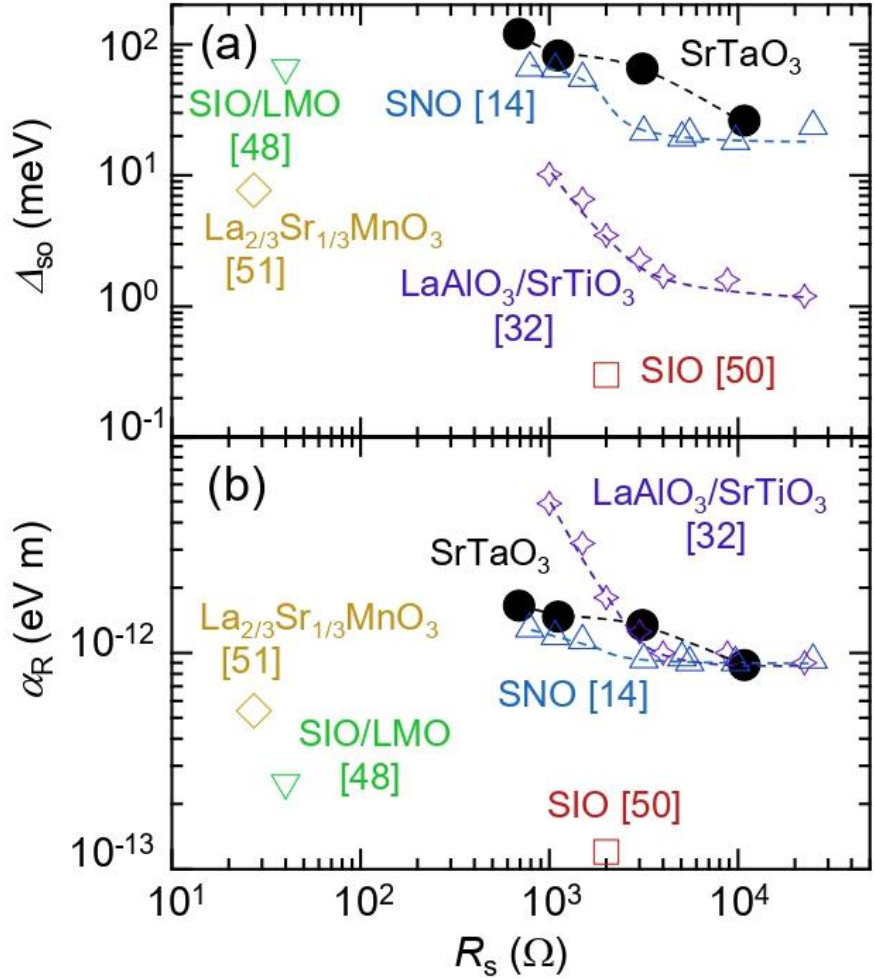

FIG. 4. (a) Energy band splitting due to SOC ($\Delta_{so}$) and (b) Rashba parameter ($\alpha_R$) for oxide materials, including SrTaO$_3$ (this study), SNO[14], SIO[50], SIO/LMO[48], La$_{2/3}$Sr$_{1/3}$MnO$_3$[51], and LaAlO$_3$/SrTiO$_3$[32] plotted against $R_s$.

In summary, epitaxial thin films of SrTaO$_3$ with $t$ smaller than 74 nm were successfully fabricated on the insulator LSAT substrate by pulsed laser deposition. Transport properties of a series of films with $t$ ranging 5.8-74 nm were measured on both sides of MIT that occurred between 8.6 nm and 5.8 nm. Above a threshold $t(\approx 25$ nm), the SrTaO$_3$ films include polycrystalline or amorphous phases, suggesting the instability of SrTaO$_3$ in thick films, and the mobility was also decreased. For the metallic films thinner than 25 nm, $L_{so}$ was deduced from the analysis of MR using the WAL theory. The values of $L_{so}$ turned out to be close to or slightly smaller than the values reported for other metallic oxide films, such as SNO and SIO thin films. The values of $\Delta_{so}$ ranged from 26 to 120 meV, which were the largest among the oxide materials. The values of $\alpha_R$ ranged from 8.8×10$^{-13}$ to 1.7×10$^{-12}$ eV m, which were much larger than the values reported for other thin films of metallic oxides, indicating that SOC in the SrTaO$_3$ thin films was the largest among the metallic oxide films reported so far.

ACKNOWLEDGMENTS

This work was supported by JSPS KAKENHI Grants No. 21H01038, No. 22K14001, No. 23KJ0604 and JST CREST Grant No. JPMJCR15Q2.

# Supplemental Material for
# "Large Rashba spin-orbit coupling in metallic SrTaO$_3$ thin films"

Hikaru Okuma[1]*, Yumiko Katayama[1], Fukunobu Kadowaki[1], Yuki Tokumoto[2], and Kazunori Ueno[1]

1 *Graduate School of Arts and Sciences, The University of Tokyo, Komaba, Meguro-ku, Tokyo 153-8902, Japan.*

2 *Institute of Industrial Science, The University of Tokyo, Komaba, Meguro-ku, Tokyo 153-8505, Japan.*

*Author to whom correspondence should be addressed: hikaruokuma613@gmail.com

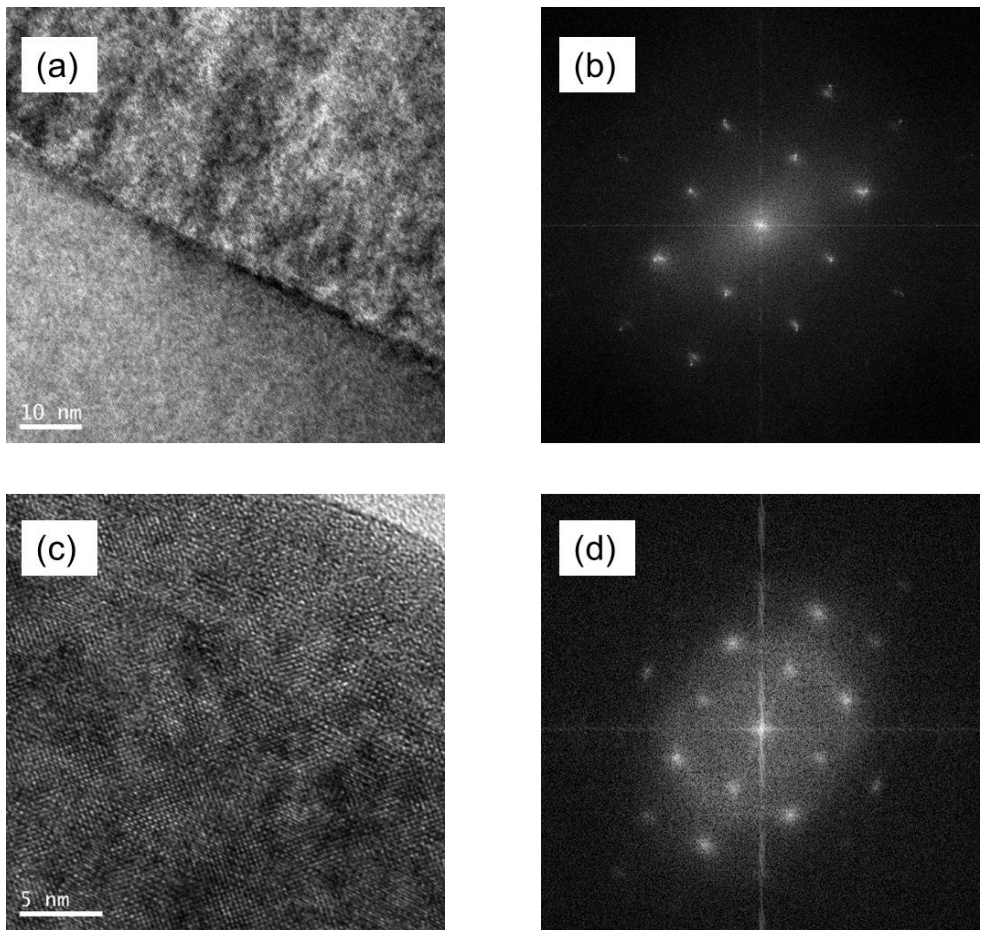

FIG. S1. (a,c) Cross-sectional TEM images for a SrTaO$_3$/LSAT film with $t$ = 74 nm (a) around the interface between the substrate and the film, and (c) around the surface. (b,d) FFT patterns of the TEM images (b) around the interface between the substrate and the film, and (d) around the surface.

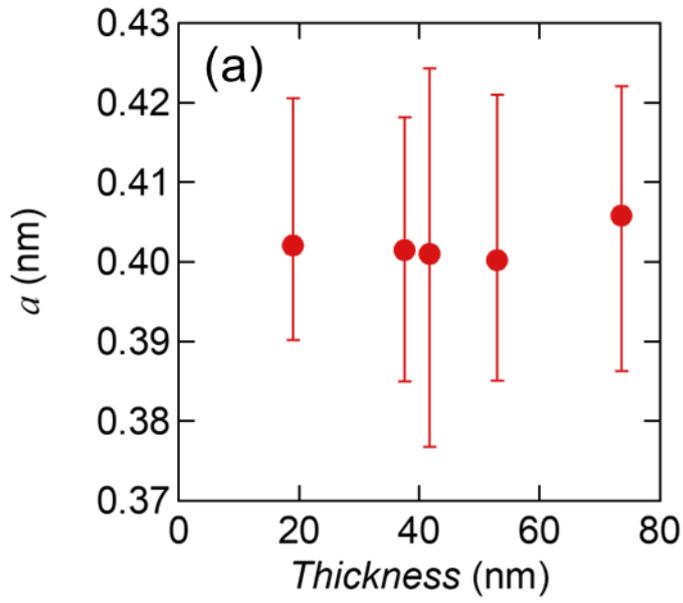
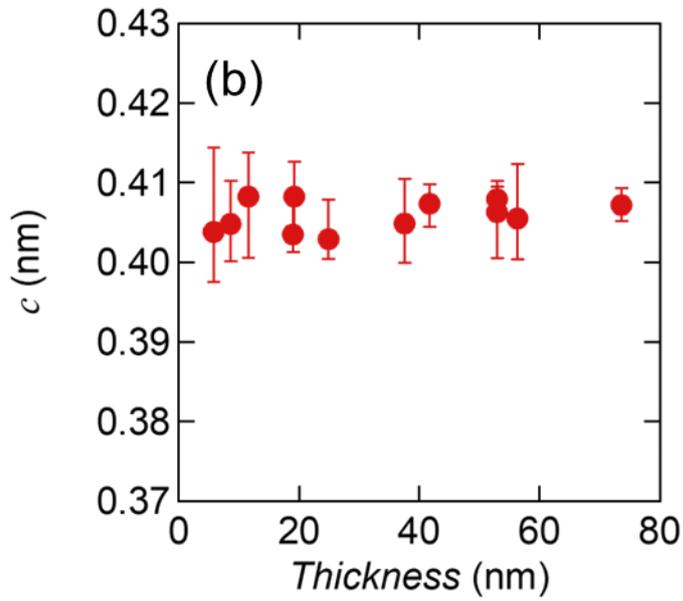

FIG. S2. (a) The in-plane ($a$) and (b) out-of-plane lattice constants ($c$) versus $t$ plots extracted from RSM and $2\theta/\omega$ scan, respectively. Error bars are defined as the full width at half maximum of the SrTaO$_3$ peak.

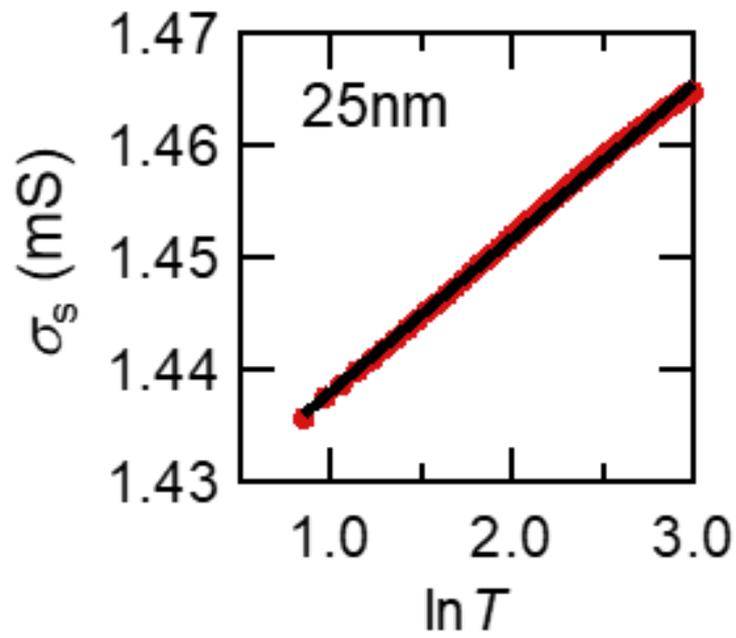

FIG. S3. Sheet conductance ($\sigma_s = 1/R_s$) versus the logarithm of temperature (ln$T$) plots for the 25 nm-thick film. The straight line represents a fit to the 2D model for the quantum interference effect.

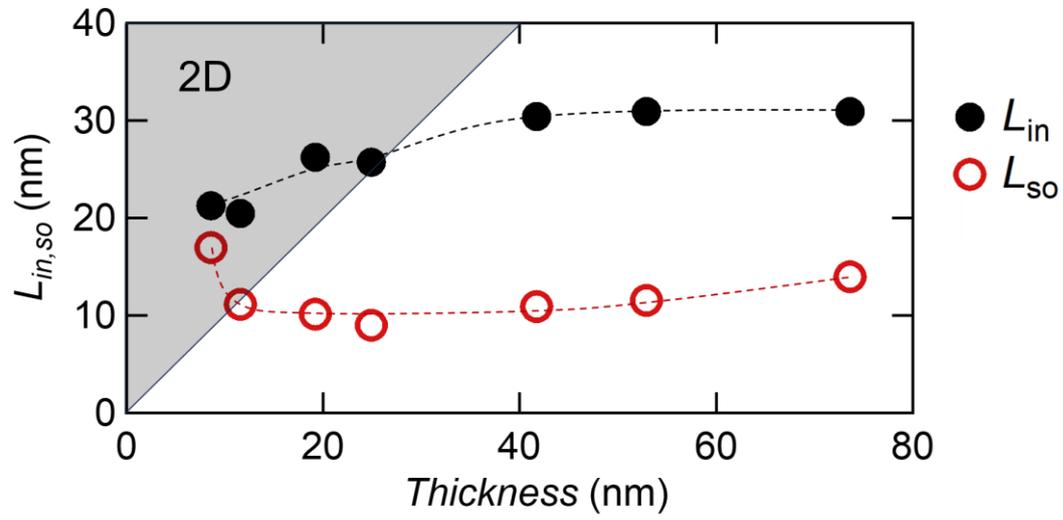

FIG. S4. Inelastic ($L_{in}$) and spin relaxation lengths ($L_{so}$) plotted against $t$. Gray shading represents the area where $L_{in}$ is larger than $t$, i.e., the 2D regime.